# A Survey on Detection of Sinkhole Attack in Wireless Sensor Network

George W. Kibirige
Department of Informatics
Sokoine University of Agriculture, SUA
Morogoro, Tanzania
georgekibirige@yahoo.com

Camilius Sanga,
Department of Informatics
Sokoine University of Agriculture, SUA
Morogoro, Tanzania
csanga@gmail.com

**Abstract**
**Wireless Sensor Network (WSN) consists of large number of low-cost, resource-constrained sensor nodes. The constraints of the wireless sensor node is their characteristics which include low memory, low computation power, they are deployed in hostile area and left unattended, small range of communication capability and low energy capabilities. Base on those characteristics makes this network vulnerable to several attacks, such as sinkhole attack. Sinkhole attack is a type of attack were compromised node tries to attract network traffic by advertise its fake routing update. One of the impacts of sinkhole attack is that, it can be used to launch other attacks like selective forwarding attack, acknowledge spoofing attack and drops or altered routing information. It can also used to send bogus information to base station. This paper is focus on exploring and analyzing the existing solutions which used to detect and identify sinkhole attack in wireless sensor network. The analysis is based on advantages and limitation of the proposed solutions.**

## I. INTRODUCTION

Wireless sensor network consists of small nodes with ability to sense and send data to base station [5]. Wireless sensor network is used in different applications example in military activities, which used to track movement of their enemy. It also used in fire detection and in healthy service for monitoring heart beat [2, 17, 13]. Unfortunately most of wireless network are deployed in unfriendly area and normally left unattended. Also most of their routing protocols do not consider security aspect due to resource constraints which include low computational power, low memory, low power supply and low communication range [8,9]. This constraint creates chance for several attackers to easily attack wireless sensor network. An example of attack is sinkhole attack. Sinkhole attack is implemented in network layer where an adversary tries to attract many traffic with the aim to prevent base station from receiving a complete sensing data from nodes [20].The adversary normally compromises the node and that node will be used to launch an attack. The compromised node send fake information to neighboring nodes about its link quality which used in routing metric to select best route during data transmission. Then all the packets from his neighbors pass through him before reach to base station. [22]. Sinkhole attack prevents base station from acquiring a complete and correct sensing data from nodes.

The purpose of this paper is to study existing solutions used to detect sinkhole attack. Different solutions which were used to detect and identified sinkhole attack were suggested by different researchers, such as Krontiris [14], Ngai et al [18] and Sheela et al [25]. Rule based detection solution were proposed by Krontiris et al[15] to detect sinkhole attack. All the rules were focused on node impersonation and were implanted in intrusion detection system. Then intruder was easily detected when they violate either of the rules. Another centralized solution which involve base station in detection process proposed by Ngai et al [18] A non cryptography scheme which used mobile agent in the network to prevent sinkhole attack was also proposed by Sheela et al [25]

The remainder of this paper is organized as follow. Section 2 discusses sinkhole attack and their attack mechanism in two different protocols. Section 3 presents the challenges in detection of sinkhole attack in wireless sensor network. Section 4 presents different approaches that proposed by different researchers to detect sinkhole attack. Finally, section 5 conclude this paper and proposed some future works.

## II. SINKHOLE ATTACK

Sinkhole attack is an insider attack were an intruder compromise a node inside the network and launches an attack. Then the compromise node try to attract all the traffic from neighbor nodes based on the routing metric that used in routing protocol. When it managed to achieve that, it will launch an attack. Due to communication pattern of wireless sensor network of many to one communication where each node send data to base station, makes this WSN vulnerable to sinkhole attack (Ngai et al [18]).

The following subsections discuss the techniques use in MintRoute protocol and AODV protocol in launching sinkhole attack.

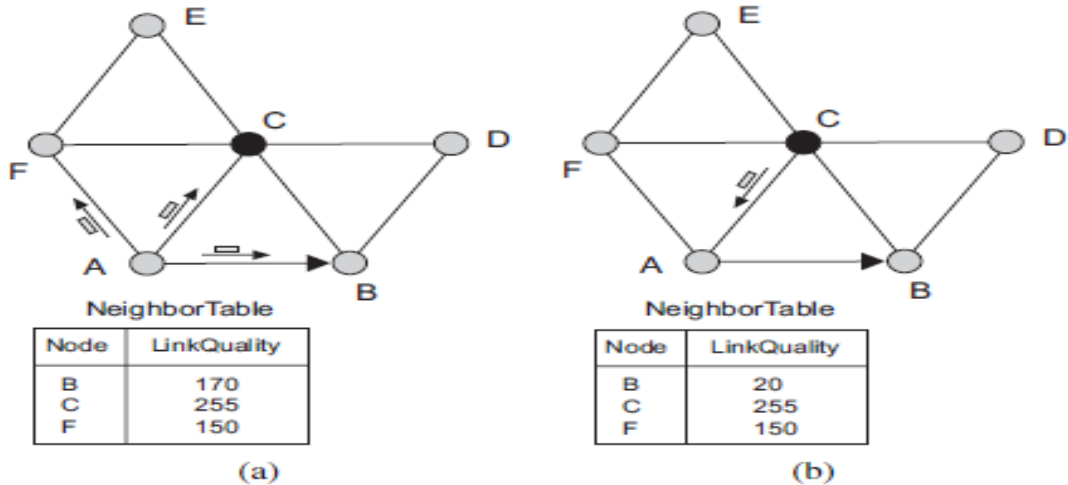

**Figure 1:** Sinkhole attack in MintRoute protocol (Krontiris, I[15])

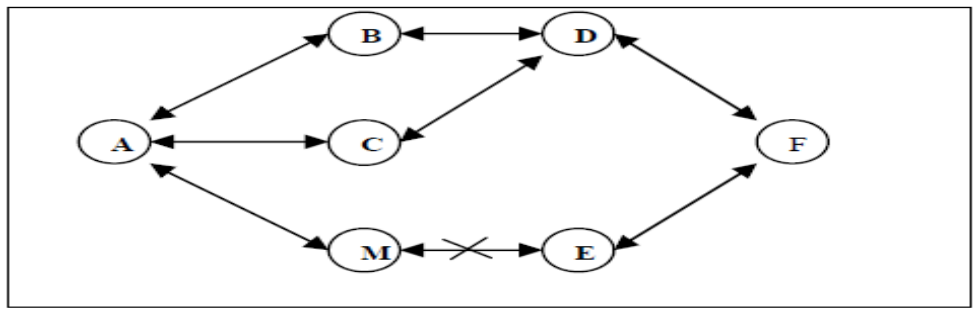

**Figure 2**: Sinkhole in TinyAODV protocol (Teng and Zhang,[27])

*Sinkhole Attack in MintRoute Protocol*

MintRoute protocol is a type of protocol which is commonly used in wireless sensor network. It was designed purposely for the wireless sensor network, it is light and suitable for sensor nodes which have minimum storage capacity, low computation power and limited power supply. MintRoute protocol uses link quality as a metric to choose the best route to send packet to the Base Station (Krontiris et al [15]).

Fig.1 shows six sensor nodes A, B, C, D, E, and F. Node C is malicious, and it is going to launch a sinkhole attack. The Figure 1(a) shows a route table of node A with IDs of its neighbors with their corresponding link quality. Originally the parent node was node B but node C advertises its link quality with a value of 255 which is maximum value. Node A is not going to change its parent node until the node B's link quality fall to 25 below the absolute value.

In Fig.1(b) the malicious node is sending new update route packet that the link quality fall up to 20 and impersonate node B so that node A believe the packet come from node B. Node A will update its route table and change the parent node to node C (Krontiris et al [15]). The attacker uses node impersonation to launch an attack.

*Sinkhole Attack in TinyAODV Protocol*
This is another explanation of sinkhole attack in wireless sensor network and this time the attack is launched under TinyAODV (Ad-hoc On Demand Vector) protocol. TinyAODV protocol is the same as AODV in MANET but this one is lighter compared to AODV and it was modified purposely for wireless sensor network [27]. The number of hops to base station is the routing metric that used in this protocol. Generally the route from source to destination is created when one of the nodes send a request, the source node sends a RREQ (Route request) packet to his neighbors when wants to send packet. Next one of the neighbors close to destination is reply by sending back RREP (Route Reply) packet, if not the packet is forwarded to other nodes close to that destination. Finally, the source receives RREP packet from neighbor then select one node with less number of hops to destination.

The sinkhole node or compromised node launches an attack by send back RREP packet. In RREP packet it gives small number of hops which indicates close proximity to the base station. Then the source node decides to forward packet to sinkhole node. The compromised node then performs the same technique to its entire neighbors and tries to attract as much traffic as possible [27].

For instance, Fig.2 shows node M launches sinkhole attack in Tiny AODV. Node A sends RREQ to nodes BCM. However node M instead of broadcast to node E like nodes B and C does to node D, he replies back RREP to node A. Then node A will reject node B and C, then forward packet to M because node A and B are very far to F compare to node M.

## III. CHALLENGES IN DETECTION OF SINKHOLE ATTACK IN WSNs

Based on the literature review of sinkhole attack in wireless sensor network, the following are the main challenges in detecting sinkhole attack in wireless sensor network

### A. Communication Pattern in WSN;

All the messages from sensor nodes in wireless sensor network are destined to base station. This created opportunity for sinkhole to launch an attack. Sinkhole attacks normally occur when compromised node send fake routing information to other nodes in the network with aim of attracting as many traffic as possible. Based on that communication pattern the intruder will only compromised the nodes which are close to base station instead of targeting all nodes in the network. This is considered as challenges because the communication pattern itself provides opportunity for attack.

### B. Sinkhole attack is unpredictable;

In wireless sensor network the packet are transmitted based on routing metric that used by different routing protocols [26]. The compromised node used its routing metric that used by routing protocol to lie to his neighbors in order to launch sinkhole attack. Then all the data from his neighbors to base station will pass through compromised node. For example the techniques used by compromised node in network that used TinyAODV protocol is different to the one used another protocol like MintRoute protocol. In MintRoute they used link quality as route metric while in Tiny AODV they used number of hop to base station as routing metric. Therefore the sinkhole attack techniques is changed based on routing metric of routing protocol

### C. Insider Attack

Insider attack and outsider attack are two categories of attack in wireless sensor network. Outside attack is when intruder is not part of network. In inside attack the intruder compromises one of the legitimate node through node tempering or through weakness in its system software then compromised node inject false information in network after listen to secret information. Inside attack can disrupt the network by modifying routing packet. Through compromised node sinkhole attack attract nearly all the traffic from particular area after making that compromised node attractive to other nodes. The fact is that compromised node possesses adequate access privilege in the network and has knowledge pertaining to valuable information about the network topology this created challenges in detecting. Base to that situation even cryptographic cannot defend against insider attack although it provides integrity, confidentiality and authentication (Pathan, K [22]). Therefore the internal attack has more serious impact on victim system compared to outsider attack.

### D. Resource Constraints;

The limited power supply, low communication range, low memory capacity and low computational power are the main constrained in wireless sensor network that hinder implementation of strong security mechanism. For example the strong cryptographic method that used in other network cannot be implemented in this network due to low computational power and low memory capacity. Therefore less strong key are considered which is compatible with available resources.

### E. Physical attack;

A wireless sensor network normally deployed in hostile environment and left unattended. This provides a opportunity for an intruder to attack a node physically and get access to all necessary information [12].

## IV. EXISTING APPROACHES

Many researchers have been working on wireless sensor field to provide security mechanism to suits the resource constrained due to growing demand of applications in sensitive areas. The following are the identified approaches that used by different researchers to detect and identified sinkhole attack in wireless sensor network. Those approaches are classified into rules based, key management, anomaly based, statistical method and hybrid based. The subsequent subsections described each of these categories and give examples of existing work that used that approach.

## A. Rule based

The rules are designed based on the behavior or technique used to launch sinkhole attack. Then those rules are imbedding in intrusion detection system which runs on each sensor nodes. Those rules were then applied to the packet transmitted through the network nodes. If any node violates the rules is considered as adversary and isolated from the network.

Among the existing work which used rules based approach include Krontiris et al [14]. Krontiris used rule based approach to detect sinkhole attack. They create two rules and implanted in Intrusion detection system (IDS). When one of the rules is violated by one of the nodes, the intrusion detection system triggered an alarm but it does not provide node ID of compromised node. The first rule "for each overhead route update packet the ID of the sender must be different your node ID". The second rule "for each overhead route update packet the ID of the sender must be one of the node ID in your neighbors". Also Krontiris et al [15] used the same approaches. There are two rules, the first rule "rule for each overhead route update packet the ID of the sender must be one of node ID in your neighbors". The second rule "for each pair of parent and child node their link quality they advertise for the link between them, the difference cannot exceed 50.

## B. Anomaly-based detection

In anomaly based detection the normal user behavior is defined and intrusion detection is searching for anything that is anomalous in the network. In this method intrusion is considered as anomalous activity because it looks abnormal compare to normal behavior. The rule based and statistical approaches are also included under anomaly based detection approach.

Tumrongwittayapak and Varakulsiripunth [29] proposed system that used RSSI (Received Signal Strength Indicator) value with the help of EM (Extra Monitor) nodes to detect sinkhole attack. The EM had high communication range and one of their functions is to calculate RSSI of node and send to base station with ID of source and next hop. This process happens instantly when node are deployed. Base station uses that RSSI value to calculate VGM (visual geographical map). That VGM shows the position of each node, then later when EM send updated RSSI value and base station identify there is change in packet flow from previous data this indicate there is sinkhole attack. The compromised node is identified and isolated from the network by base station using VGM value. However, if attack is launched immediately after network deployment, the system will not be able to detect that attack [29]. Also the numbers of EM nodes were not specified for specific number of sensor nodes and the proposed method is focused only on static network.

## C. Statistical method

In statistical approaches the data associated with certain activities of the nodes in network is studied and recorded by researchers. For example monitor the normal packet transmitted between the nodes or monitor resource depletion of the nodes like CPU usage. Then the adversary or compromised node is detected by comparing the actual behavior with the threshold value which used as reference, if any nodes exceed that value is considered as an intruder.

Chen, et al [3], proposed statistical GRSh (Girshick-RubinShyriaev)–based algorithm for detecting malicious nodes in wireless sensor network. Base station calculates the difference of CPU usage of each node after monitoring the CPU usage of each node in fixed time. Base station would identify whether a node is malicious or not after comparing the difference of CPU usage with the threshold.

Dynamic trust management system was proposed by Roy et al [23] to detect and eliminate multiple attacks such as sinkhole attack. Each node calculates the trust of its neighbor node based on experience of interaction; recommendation and knowledge then sends to base station. The base station decided which node is sinkhole after it received several trust values from other nodes. Therefore the trust value of the node which falls beyond the normal value 0.5 is considered as sinkhole attack [23].

## D. Hybrid based intrusion detection

The combination of both anomaly and signature based or misused based is used in this approach. The false positive rate which produced by anomaly based is reduced in this approach due to the use of both method. Also the advantage of this approach is to be able to catch any suspicious nodes which their signature is not included in detection database.

Coppolino and Spagnuolo [6] proposed hybrid Intrusion detection system to detect sinkhole attack and other attacks. They used detection agent which was responsible for identifying sinkhole attack. The hybrid intrusion detection was attached to sensor node and share resource of that node. The suspicious nodes were inserted to the blacklist based on anomalous behavior after analyzed the collected data from neighbors. Then that list is sent to central agent to make final decision based on feature of attack pattern (misused based). Similar to solution proposed by Tumrongwittayapak and

Varakulsiripunth [29], it was designed for static wireless sensor network.

*E. Key management*

In key management approach the integrity and authenticity of packet travels within the network is protected by using encryption and decryption key. Any packet transmitted in the network is added with another message in a way that to access that message requires a key and any small modification of the message can be easily detected. Those keys also help nodes to check if the message comes from base station and check the authenticity of the message.

Papadimitriou et al [21] proposed a cryptographic approach in routing protocol to address the problem of sinkhole attack. Each node obtained public key which used to verify if the message comes from base station. They also used pair of public and private keys for authentication and sign data message. All keys were uploaded offline before the network was deployed. Their techniques prevented any node to hide its ID and any packet forgery between nodes in the network. This protocol is focused on resistance to sinkhole attack but not to detect and eliminate it.

Meanwhile, Fessant et al [10] proposed two protocols which used cryptographic method to increase the resilience of sinkhole attack. Both protocols prevent malicious node from lying about their advertised distances to base station. However, they did not show the memory usage of their protocols and message size.

The summary of existing works using the previously described approaches is shown in Table 1. The summary covers evaluation results of proposed solution and their limitations

**Table 1: Existing works on Sinkhole detection**

| Approach | Proposed Solution | Result | Limitations/Advantages |
|---|---|---|---|
| Rule Based. Krontiris et al 2007 [16] | They extended their IDS which can detect sinkhole attack. | - the success of intrusion detection system depend on the increase number of watchdog<br>- When the network density increase the false negative rate decrease. | Limitations<br>- Memory and network overhead was created.<br>- They used MintRoute protocol<br>- Node impersonation was the focus of the rules.<br><br>Advantages<br>- More secure and robust measure can be developed based on valuable principle they develop. |
| Rule Based. Krontiris et al 2008 [15] | They proposed detection rules that will keep aware legitimate node the existing of attack. | - They show how vulnerabilities of MultihopLQI can be exploited by sinkhole node and suggest the rules which make the protocol more resilient. | Limitation<br>- They did not show practically how those rules can prevent attack.<br>- All the rules are only detecting attack but cannot give ID of sinkhole node.<br>- They assume attacker has the same power as normal node and can capture sensor node and change the internal state. |
| Anomaly based. Tumrongwittayapak, C and Varakulsiripunth, R 2009 [29] | They proposed detection solution based on received signal strength indicator(RSSI)<br><br>Their proposed | - For 0 to 40% percentage of message drop the detection rate is 100%<br>- False positive rate was 0 for 0-40% of message drop but increase when percentage drop increase<br>- The same applied to false negative rate with the more | Limitation<br>- They assume sensor network are static<br>- No instant attack<br>- Base station remain 0,0 position<br>- Base station and extra monitor node are physically protected.<br>- Their proposed solution can not detect attack if it happened |

| | solution required support from extra monitor node | message drop the more negative rate. | instantly after network deployment. |
|---|---|---|---|
| Anomaly based. Choi et al 2009 [4] | They proposed method that can detect sinkhole attack that used LQI (link quality indicator). | <ul><li>The probability of detection increase when number of detector nodes increase</li><li>detection rate increase when detector node increase</li><li>The false positive rate depend on extent of tolerance value (constant value which will show if changes is beyond abnormal)</li></ul> | Limitations<ul><li>All sensor node have no mobility</li><li>The detection of sinkhole occurs when detector node is between sinkhole node and source node and sinkhole and base station</li><li>The detector nodes have high source of energy than sensor nodes</li></ul>Advantage<ul><li>Detector node communicate themselves through exclusive channel</li></ul> |
| Anomaly based. Sharmila, S. and Umamaheswari, G. 2011. [24] | -They proposed message digest algorithm to detect sinkhole node. | <ul><li>The results show the algorithm worked well when malicious nodes are below 50%</li><li>False positive rate was 20%( due to packet drop) that figure obtained when malicious node reach 50</li><li>False negative error was 10% but was increasing when malicious node reach above 40</li></ul> | Limitation<ul><li>Network throughput, overhead and communication cost was not calculated</li><li>The performance was not good when there is node collision, limited transmitted power and packet drops</li><li>Only one advertisement is considered at a time, after computation another take place</li></ul>Advantage<ul><li>The algorithm achieve data integrity and authenticity</li></ul> |
| Key Management. Papadimitriou et al 2009 [21] | -They proposed two RESIST protocols which increase resilience to sinkhole attack in WSN | -Results show that RESIST-0 has high resilience to sinkhole attack (it does not allow node to lie about their distance to base station) than other protocol | Limitation<ul><li>Resist-0 is very expensive it require two additional message to a packet</li><li>In their simulation message losses and collusion were not considered</li><li>Collusion node has impact on RESIST-0 not in RESIST-1</li><li>Their routing algorithm relying on tree-based topology construction</li><li>Route tree is built by hop distance</li></ul>Advantage<ul><li>RESIST-1 prevent malicious nodes from changing their advertised distance to the sink more than one hop</li><li>RESIST-0 completely stops any</li></ul> |
| | | | lying about distance. |
| Statistical based Chen et al 2010 [3] | They develop an algorithm which detect sinkhole attack and identified intruder. | • From first simulation the detection time increase when threshold (CPU value) become bigger<br>• Also the false positive rate decrease when threshold become bigger.<br>• From the second simulation the detection time did not change too much but the false positive rate increase due to increase in traffic | Limitation<br>• Base station makes the final decision on which node is malicious<br>• No results on the network overhead<br>• The scheme will not detect attack if it launch instantly after deployed<br>• Assumption-base station is trustworthy and it participates in detection system.<br>Advantages<br>• Their algorithm showed that it can detect malicious node in short time with low false positive rate |
| Hybrid base Coppolino et al 2007 [6] | They proposed intrusion detection system which was able to protect critical information from attacks directs from its WSN. | • Detection rate was 95-97% when malicious node modified sensor packet.<br>• Detection rate was 93-96% when malicious node modified the r<br>• False positive rate is 3%<br>• IDS usage in real sensor network was 734bytes (RAM) and 3208bytes (ROM) | Advantage<br>• Their solution satisfied the available resource in sensor nodes<br>• Their solution proved to detect sinkhole attack<br>• They used both anomaly and misuse based method |
| A non cryptographic Sheela, D et al 2011 [25] | They proposed scheme which used mobile agent to defend against this attack | • Probability of detecting sinkhole is decrease when nodes increase<br>• Node average energy decrease as time goes up because of storage information.<br>• The algorithm create high network overheads | Limitation<br>• Mobile wireless sensor network<br>• No specification of exactly number of MA(mobile agent) in network<br>• Matrix method is very complex with relate to available resources<br>• MA communicate with sensor nodes at active mode only<br>Advantage<br>• MA used dummy data to detect modification<br>• MA has sufficient power to run its activities |

## V. DISCUSSION

From the Table 1, it shows most approaches managed to detect and prevent sinkhole attack in WSN.

Rule based approaches managed to detect sinkhole attack but it creates memory and network overhead. This approach did not give the ID of sinkhole node after detection of attack. All the rules focus on the node impersonation.

Anomaly based approach also manage to detect sinkhole attack but they just focus on static wireless sensor network. This approach created high false positive rate when there was high message dropping.

Key management was another approach which focused on resistance to sinkhole attack but not to detect and eliminate it.

Statistical based approach managed to detect sinkhole attack but they did not give result of the network overhead. Also this approach cannot detect an instant attack after WSN is deployed. False positive rate were the main drawback to this approach.

Hybrid based intrusion detection approach used the combination of both anomaly and signature-based. This approach detected sinkhole attack but was designed for static WSN. It produced less false positive rate.

A non cryptographic is another approach which detected sinkhole attack but it created high network overhead.

All the approaches managed to detect, identify and provided resistance to sinkhole attack. The major drawbacks produce by those approaches includes high network and memory overhead, create high false positive rate and some were not able to work on mobile WSN.

## VI. CONCLUSION AND FUTURE WORK

Based on existing works most researchers are trying to look for ICT solutions for detecting, identifying and providing resistance to sinkhole attack in wireless sensor network. Researchers used intrusion detection scheme based on anomaly-method, other used rule based and key management to detect and identifying the sinkhole nodes. Majority of researches struggled with security challenges corresponding with availability of resources and mobility of wireless sensor nodes. Some provided solution for only static and few on mobile network. Very few researchers managed to validate their security system using real wireless sensor network. Also some of results showed low detection rate, high network overhead and high communication cost. The future solution should focus on reducing high network overhead, computational power, increase detection rate and that system must be validated in real sensor network. Through this kind of validation, it will be easy to check if their solutions meet the available resources of WSN, such as memory capacity.

*ACKNOWLEDGEMENT*

We would like to thank almighty God for giving us strength during the all period of the research. We would to like to express our sincere appreciation to other colleague in our department for their advice.

*REFERENCES*

[1]. Abhishek Pandey and R.C. Tripathi. (2010). A Survey on Wireless Sensor Networks Security, International Journal of Computer Applications (0975 – 8887) Volume 3 – No.2.

[2]. Changlong Chen, Min Song, and George Hsieh (2010) Intrusion detection of Sinkhole attack in large scale Wireless Sensor Networks, In Wireless Communications, Networking and Information Security (WCNIS), 2010 IEEE International Conference on (pp.711-716).IEEE

[3]. Chen, C., Song, M. and Hsieh, G. (2010). Intrusion Detection sinkhole attack in large scale wireless sensor network, In Wireless Communication, Networking and Information Security (WCNIS), 2010 IEEE Interational Conference on (pp. 711-716). IEEE.

[4]. Choi, G. B., Cho, J. E., Kim,H. J., Hong, S. C. and Kim, H. J. (2008).A sinkhole attack detection mechanism for LQI based mesh routing in WSN. In ICOIN (pp.1-5).

[5] Chun-ming Rong , Skjalg Eggen, Hong-bing Cheng. (2011). A Novel Intrusion Detection Algorithm for Wireless Sensor networks. In Wireless Communication, Vehicular Technology, information Theory and Aerospace & Electronic Systems Technology (Wireless VITAE), 2011 2$^{nd}$ International Conference on (pp. 1-7).

[6] Coppolino, L., D'Antonio, S., Romano, L., and Spagnuolo, G.(2010). An intrusion detection system for critical information infrastructures using WSN technologies. In Critical Infrastructure (CRIS), 2010 5$^{th}$ International Conference on (pp. 1-8). IEEE

[7] Daniel Dallas, Christopher Leckie, Kotagiri Ramamohanarao. (2007). Hop-Count Monitoring: Detecting Sinkhole Attacks in Wireless Sensor Networks, In Networks, 2007. ICON 2007. 15$^{th}$ IEEE International Conference on (pp. 176-181)

[8]. David Martins and Hervé Guyennet. (2010) Wireless Sensor Network Attacks and Security Mechanisms: A Short Survey, In Network-Based Information Systems (NBiS), 2010 13$^{th}$ International Conference on (pp. 313-320). IEEE

[9]. Edith C. H. Ngai, Jiangchuan Liu and Michael R. Lyu. (2006). On the Intruder Detection for Sinkhole Attack in Wireless Sensor Networks, In Communications, 2006. ICC'06. IEEE International Conference on (Vol.8, pp. 3383-3389). IEEE.

[10].Fessant, F., Papadimitriou, A., Viana, A.,Sengul, C. and Polamar, E. (2011) A sinkhole resilient protocol for wireless sensor network: Performance and security analysis. Computer Communications, 35(2), 234-248.

[11]. G.H. Raghunandan, B.N. Lakshmi. (2011.). A Comparative Analysis of Routing Techniques for Wireless Sensor Networks, In Innovations in Emerging Technology (NCOIET), 2011 National Conference on (pp. 17-22). IEEE

[12]. Jaydip Sen. (2009). A Survey on Wireless Sensor Network Security, International Journal of Communication Networks & Information Security, 1(2).

[13] Kalpana Sharma and M K Ghose. (2010) Wireless Sensor Networks: An Overview on its Security Threats, IJCA, Special Issue on "Mobile Ad-Hoc Networks" MANET.

[14]. Krontiris,I., Dimitriou,T., Giannetsos,T. and Mpasoukos, M. (2008). Intrusion Detection Sinkhole Attacks in Wireless Sensor Network. In Networking and Communications, 2008. WIMOB'08. IEEE Interational

Conference on Wireless and Mobile Computing, (pp. 526-531). IEEE.

[15]. Krontiris, I., Giannetsos, T. and Dimitriou, T. (2008). Launch Sinkhole Attack in Wireless Sensor Network; the Intruder Side. In Networking and Communications, 2008. WIMOB'08. IEEE International Conference on Wireless and Mobile Computing, (pp. 526-531). IEEE.

[16] Krontiris, I. Dimitrou, T. Freiling, F.C. (2007). Towards intrusion detection in wireless sensor networks. In Proc. Of the 13[th] European Wireless Conference.

[17]. Liping Teng and Yongping Zhang. (2010). SeRA: A Secure Routing Algorithm against Sinkhole Attacks for Mobile Wireless Sensor Networks, In Computer Modelling and Simulation, 2010. ICCMS'10. Second International Conference on (Vol. 4, pp. 79-82).

[18]. Ngai, E., Liu, J and Lyu, M. (2007) An efficient intruder detection algorithm against sinkhole attack in wireless sensor network. Computer Communications, 30(11), 2353-2364.

[19]. Ngai, E., Liu, J. and Lyu, M. (2006). On intruder detection for Sinkhole attack in Wireless Sensor Network. In Communications, 2006. ICC'06.IEEE International Conference on (Vol. 8, pp.3383-3389). IEEE

[20]. P. Samundiswary, D.Sathian and P. Dananjayan. (2010). Secured greedy perimeter stateless routing for wireless sensor networks, International Journal of Ad hoc, Sensor & Ubiquitous Computing( IJASUC )1, (2)

[21]. Papadimitriou, A., Fessant, L. F. and Sengul, C. (2009). Cryptographic protocols to fight sinkhole attacks on tree based routing in WSN. In Secure Network Protocols, 2009. NPSec 2009. 5[th] IEEE Workshop on (pp.43-48). IEEE

[22]. Pathan, K., Al-S. (2011) Security of Self-Organizing Networks-MANET, WSN, VANET, WMN. ISB N-13:978-1-4398-1920-3. Taylor and Francis Group.

[23] Roy, D.S., Singh, A.S. and Choudhury, S. (2008). Countering Sinkhole and Blackhole Attacks on Sensor Networks using Dynamic Trust Management. In Computers and Communications, 2008. ISCC 2008. IEE Symposium on (pp. 537-542). IEEE.

[24]. Sharmila, S. and Umamaheswari, G. (2011). Detection of sinkhole attack in WSN using message digest algorithms. In Process Automation, Control and Computing (PACC), 2011 International Conference on (pp. 1-6). IEEE.

[25]. Sheela, D., Kumar, N., and C Dr. Mahadevan, G.C.(2011). A non-Cryptographic Method of Sinkhole Attack Detection in Wireless Sensor Networks. In Recent Trends in Information Technology (ICRTIT), 2011 International Conference on (pp.527-532). IEEE

[26]. Suman Deb Roy, Sneha Aman Singh, Subhrabrata Choudhury, and N. C. Debnath. (2008). Countering Sinkhole and Black hole Attacks on Sensor Networks using Dynamic Trust Management", In computers and Communications, 2008. ISCC 2008. IEEE Symposium on (pp.537-542). IEEE.

[27]. Teng, L., and Zhang, Y. (2010). Secure Routing Algorithm against Sinkhole attack for Mobile Wireless Sensor Network, In Computer Modeling and Simulation, 2010. ICCMS'10. Second International Conference on (Vol. 4 pp.79-82). IEEE..

[28]. Tumrongwittayapak, C and Varakulsiripunth, R. (2009). Detecting sinkhole attack and Selective forwarding attack in WSN. Information Communications and Signal Processing, 2009. ICICS 2009, 7[th] International Conference on (pp.1-5). IEEE.

[29] Tumrongwittaya and Varakulsiripunth. (2009). Detection of Sinkhole attack in Wireless Sensor Networks, In ICCAS-SICE, 2009 (pp. 1966-1971). IEEE..

## AUTHOR PROFILE

**Dr. Camilius Sanga** is Associate Professor from the Department of Informatics at Sokoine University of Agriculture (SUA), Tanzania. He is a head of the Department of Informatics at the Faculty of Science, SUA. He has PhD in Computer Science from the University of the Western Cape, South Africa. He holds MSc. Computer Science and BSc. in Computer Science from Osmania University and University of Dar es Salaam respectively.

**Mr. George W. Kibirige** is Assistant Lecturer from the Department of Informatics at Sokoine University of Agriculture (SUA), Tanzania. He has MSc. Computer Science (Information Security) and BSc. in ICT Management from University of Technology Malaysia and Mzumbe University.